\documentclass[preprint,12pt]{elsarticle}

\usepackage{amssymb}

\journal{Acta Materialia}

\newcommand{\E}{{\rm e}}

\newcommand{\I}{{\rm i}}

\newcommand{\D}{{\rm d}}

\newcommand{\beq}[1]{
\begin{equation}
\label{e#1} }

\newcommand{\eeq}{
\end{equation}
}

\renewcommand{\:}{\ifmmode\mkern4mu\else\kern0.22222em\fi } 
\newcommand{\oC}{${\:{}^\circ\mkern-2mu\rm C}$} 

\usepackage{graphicx}
\usepackage{graphics}
\usepackage{color}
\usepackage{bm}

\begin{document}

\begin{frontmatter}



\title{Ordered array of $\omega$ particles in $\beta$-Ti matrix studied by small-angle x-ray scattering}


\author{J. \v{S}milauerov\'{a}\corref{cor1}}
\cortext[cor1]{Corresponding author}
\ead{jana.smilauerova@gmail.com}

\author{P. Harcuba}
\ead{Petr.Harcuba@mff.cuni.cz}

\author{J. Str\'{a}sk\'{y}}
\ead{josef.strasky@gmail.com}

\author{J. Str\'{a}sk\'{a}}
\ead{straska.jitka@gmail.com}

\author{M. Jane\v{c}ek}
\ead{janecek@met.mff.cuni.cz}

\author{J. Posp\'{\i}\v{s}il}
\ead{jiri.pospisil@centrum.cz}

\author{R. Ku\v{z}el}
\ead{kuzel@karlov.mff.cuni.cz}

\author{T. Brun\'{a}tov\'{a}}
\ead{brunatovat@centrum.cz}

\author{V. Hol\'{y}}
\ead{holy@mag.mff.cuni.cz}
\address{Faculty of Mathematics and Physics, Charles University in Prague, Czech Republic}

\author{J. Ilavsk\'{y}}
\ead{ilavsky@aps.anl.gov}
\address{Argonne Nat. Laboratory, Argonne Ill., USA}

\begin{abstract}
Nano-sized particles of $\omega$ phase in a $\beta$-Ti alloy were
investigated by small-angle x-ray scattering using synchrotron
radiation. We demonstrated that the particles are spontaneously
weakly ordered in a three-dimensional cubic array along the
$\langle 100 \rangle$-directions in the $\beta$-Ti matrix. The
small-angle scattering data fit well to a three-dimensional
short-range-order model; from the fit we determined the evolution
of the mean particle size and mean distance between particles
during ageing. The self-ordering of the particles is explained by
elastic interaction between the particles, since the relative
positions of the particles coincide with local minima of the
interaction energy. We performed numerical Monte-Carlo simulation
of the particle ordering and we obtained a good agreement with the
experimental data.
\end{abstract}


\end{frontmatter}


\section{Introduction}
\label{introduction}

Metastable $\beta$-Ti alloys are increasingly used in aerospace
and automotive industry mainly due to excellent corrosion
resistance and high specific strength. The high strength is achieved
through ageing treatment involving several phase transformations
\cite{Banerjee2013}. Therefore, investigation of these phase
transformations is of significant importance.

Above 883\oC{}, pure titanium crystallizes in a body-centered cubic
structure ($\beta$ phase). When cooled below this temperature
($\beta$-transus) it martensitically transforms to a hexagonal
close-packed structure ($\alpha$ phase). Metastable $\beta$-Ti
alloys contain a sufficient amount of $\beta$-stabilizing elements
(Mo, V, Nb, Fe) so that the martensitic $\beta \rightarrow \alpha$
transformation is suppressed and the $\beta$ phase is retained
after quenching to room temperature \cite{lutjering2007titanium}.

Several metastable phases can emerge during ageing of these alloys
depending on the content of $\beta$-stabilizing elements.
The present study focuses on hexagonal $\omega$ phase. Tiny and
uniformly distributed particles of the $\omega$ phase serve as
precursors for a subsequent precipitation of the $\alpha$-phase
particles that are responsible for significant strengthening.

The $\omega$ phase is formed upon quenching by a diffusionless
displacive transformation as first proposed by Hatt et al.
\cite{hatt1960omega} and lucidly described by de~Fontaine
\cite{defontaine1970mechanical}. The transformation can be
described as a collapse of two neighbouring (111)$_\beta$ planes
into one plane. More formally, these two planes are displaced by
$\pm \frac{1}{12} \left[111 \right]$ along the body diagonal of
the cubic unit cell. One (111)$_\beta$ plane between two pairs of
collapsed planes remains unchanged. This produces a hexagonal
structure with two differently populated alternating 'basal'
planes. Such hexagonal structure is coherent with the parent
$\beta$ phase \cite{williams1973omega}. It was shown
experimentally that this displacive $\beta \rightarrow \omega$
transformation is completely reversible at low temperatures at
which diffusion does not play a role \cite{defontaine1971omega}.
The $\omega$ phase forms fine, a few nanometers large particles
uniformly dispersed throughout the $\beta$ matrix. Due to its
formation mechanism, the $\omega$ phase can exist only in certain
orientations with respect to the $\beta$ matrix. The topotactical
relationship between the $\beta$ and $\omega$ lattices can be
described as \cite{silcock1958x}
\beq{10}
(0001)_\omega \parallel (111)_\beta,\
\left[11\bar{2}0\right]{}_\omega \parallel [011]_\beta.
\eeq
The particles of the $\omega$ phase further evolve and grow during
ageing through a diffusion controlled reaction \cite{devaraj2009}.
This process is irreversible and is accompanied by rejection of
$\beta$ stabilizing elements from the $\omega$ phase.

It has been observed that the shape of the $\omega$ particles can
be either ellipsoidal or cuboidal. According to Blackburn et al.
\cite{blackburn1968}, the shape is related to the lattice misfit
strain. They suggested that the ellipsoidal shape of the $\omega$
precipitates arises from an anisotropy of the strain energy of the
precipitate rather than the matrix, whereas the cuboidal shape is
determined by the strain energy in the matrix.

Despite countless studies describing the $\omega$ phase, there is
still not much known about the causes of formation of the $\omega$-phase particles and their spatial ordering. Some argue that the
$\omega$ particles formation follows from a spinodal chemical
separation of the $\beta$ phase \cite{koul, ng}. This conclusion is
based on observations of the spatial ordering and/or chemical
inhomogeneity of the particles and the host material. Others suggest that
the formation of these particles can be attributed to elastic
instabilities of the parent $\beta$ matrix
\cite{DevarajActaMat2012}.

In the last decades, kinetics of the phase separation in alloys
has been intensively studied both theoretically and by various
experimental methods. For the theoretical description of the phase
separation both macroscopic and microscopic approaches have been
published, the former describes the phases as elastic continua
divided by ideally sharp interfaces, the latter takes into account
movement of individual atoms. The description is based on
classical works of Cahn and Hilliard
\cite{CahnTMS1968,HilliardPhT1970} and Lifshitz, Slyozov and
Wagner \cite{LifshitzJPCS1961,WagnerZEE1961} (LSW theory) and
includes various processes denoted spinodal decomposition,
coarsening or Ostwald ripening (see also the review in
\cite{FratzlJSM1999}). From numerous numerical simulations and
experimental data it follows that the time-dependent structure
function
\beq{a1}
S(Q,t) = \left\langle \left| \int \D^{3} {\bm r} c({\bm r})
\E^{-\I {\bm Q}.{\bm r}} \right|^2 \right\rangle,
\eeq
i.e., the square of the modulus of the Fourier transformation of
the concentration function $c({\bm r})$ of a given phase averaged
over a statistical ensemble, obeys a universal scaling law
\cite{LebowitzACM1982}
\beq{a2}
S(Q,t) \sim Q_{\rm max}(t)^{-3} F(Q/Q_{\rm max}(t)),
\eeq
where $Q_{\rm max}(t)$ is the position of the maximum of the
structure function $S(Q,t)$ and $F$ is a time-independent
universal function. From the time-dependence of $Q_{\rm max}$ we
can deduce the evolution of the characteristic length $L(t) \sim
2\pi/Q_{\rm max}(t)$ during the coarsening process; from the LSW
theory the asymptotic behavior $Q_{\rm max}(t)^{-3} \sim A + Bt$
follows.

$\omega$ phase particles in metastable $\beta$-Ti alloys are
nanometers in size and after ageing have slightly different
electron density than the parent $\beta$ matrix  (i.e. the x-ray
indexes of refraction of the particles and the matrix are
different). Small-angle x-ray scattering (SAXS) is an
ideal technique to determine the structure function, since the
reciprocal-space distribution of the scattered intensity is
proportional to the structure function multiplied by the square of
the difference in the electron densities of the $\beta$ and
$\omega$ phases. SAXS is a nondestructive technique based on
elastic scattering from electron density inhomogeneities within
the sample. The SAXS instrument records scattered intensity at small
scattering angles, i.e., close to the direction of the incident beam.
Obtained SAXS data contain information about important
microstructural parameters such as size, shape, volume and space
correlations of the scatterers \cite{Ilavsky2009}.

To our knowledge, only a limited number of experiments employing
SAXS on $\omega$ particles in titanium alloys has been performed
and published. Fratzl et al. \cite{FratzlActaMet1991} investigated
the growth of $\omega$-phase particles in single crystals of
Ti--20 at.\%Mo. They found that the radii of the $\omega$ particles
increased with ageing time as $\sim t^{1/3}$ and then stabilized
at the value of approximately 75\:{}\AA. The same group of authors
later investigated $\omega$ and $\alpha$ phase precipitation in
Ti--12 at.\%Mo single crystals by the means of SAXS
\cite{LangmayrPRB1994}. In their work, the authors determined the
shape of the $\omega$ particles and then observed the nucleation and
coarsening of $\alpha$ plates which destroyed the $\omega$
structure.

In our previous paper \cite{SmilauerovaActaMat2013} we studied the
structure of the $\omega$ particles in a single-crystalline $\beta$
matrix by x-ray diffraction (XRD). We confirmed the validity of
the topotactical relations in Eq. (\ref{e10}) and found that the
$\beta$ lattice is locally compressed around the particles. In
this paper we perform a systematic SAXS study of the evolution of
sizes of the $\omega$ particles in single-crystalline $\beta$-Ti alloy
samples during ageing. We demonstrate that the $\omega$ particles
are self-ordered and create a three-dimensional cubic array along
the crystallographic axes $\langle 100 \rangle_\beta$ of the
$\beta$ matrix and that the mean distance between the particles is
proportional to their mean size. We explain the ordering mechanism
by considering the energy of the elastic interaction between the
particles caused by the local deformation of the lattice. Finally,
we compare the measured data with the scaling behavior in Eq.
(\ref{ea2}).

The paper is written as follows. The next section contains a brief
description of the growth of the Ti-alloy single crystals and the
SAXS experiments. Section \ref{SRO} contains a phenomenological
model of the arrays of particles that makes it possible to fit the
experimental data and to determine basic structural parameters.
The results of the SRO model are also presented in Section~\ref{SRO}.
The driving force of the self-ordering process is discussed in
Section \ref{drivingforce}. The Section~\ref{discussion} provides
a discussion of the obtained results.

\section{Experiments}
\label{experiments}

Single crystals of one of the metastable $\beta$ titanium alloys,
TIMETAL LCB were grown in a commercial optical floating zone
furnace (model FZ-T-4000-VPM-PC, Crystal Systems Corp., Japan)
with four 1000 W halogen lamps. The growth process was carried out
in a protective Ar atmosphere with Ar flow of 0.25~l/min and
pressure of 2.5~bar. The growth speed was 10~mm/h. The grown
ingots had roughly circular cross section with diameter in the
range of 8~--~10~mm. The length of the single crystals was
typically around 9\:{}cm. The details of the single crystal growth
process and characterization of the resulting ingot can be found
elsewhere \cite{growth}. During single crystal growth, a shift in
chemical composition of the material is possible. In order to
assess this change, concentrations of individual elements were
determined both for precursor material and grown ingot. Two
experimental methods were used. Concentrations of the main
alloying elements (Ti, Mo, Fe, Al) were determined by energy
dispersive x-ray spectroscopy (EDX) using scanning electron
microscope FEI\:{}Quanta\:{}200FEG. Because titanium alloys are
prone to contamination by interstitial N and O, the concentrations
of these elements were checked by an automatic analyzer
LECO\:{}TC\:{}500C. The chemical composition of the precursor and
resulting single crystal are summarized in Tab. \ref{t1}.

\begin{table}[!h]
\scalebox{0.87}{
\begin{tabular}{|l||l|l|l|l|l|l|}
\hline
sample  &Ti             & Mo            & Fe            &Al         &  N                & O \\ \hline \hline
precursor   & $88.7 \pm 0.7$ & $4.1 \pm 0.4$ & $3.7 \pm 0.5$ & $3.1 \pm 0.2$ & $0.011 \pm 0.004$ & $0.40 \pm 0.06$ \\
crystal & $88.4 \pm 0.4$ & $4.3 \pm 0.3$ & $3.4 \pm 0.4$ & $3.1 \pm 0.2$ & $0.29 \pm 0.03$ & $0.55 \pm 0.06$ \\
\hline
\end{tabular} }
\caption{Chemical composition of the precursor material and the
resulting single crystal in at. \%} \label{t1}
\end{table}

Each crystal was solution treated at 860\oC{} for 4\:{}h in an
evacuated quartz tube and water quenched in order to homogenize
the structure and ensure the retention of $\beta$ phase.
Subsequently, the crystals were cut into 1.2\:mm thick  slices
perpendicular to the length of the crystal. The slices were then
aged in salt bath at three temperatures 300\oC{}, 335\oC{}, and
370\oC{} for 2, 4, 8, 16, 32, 64, 128 and 256\:{}h. At these
ageing temperatures, the $\omega$ phase particles grow, but at least at the lowest temperatures in the series, the
precipitation of the $\alpha$-Ti phase is not expected
\cite{PrimaJIM2000,azimzadeh1998phase}. The samples  were then
ground and polished from both sides utilizing 500, 800, 1200, 2400
and 4000 grit SiC papers. Final polishing was carried out on a
vibratory polisher using 0.3\:{}$\mu$m and 0.05\:{}$\mu$m aqueous
alumina (Al$_2$O$_3$) suspensions and 0.05\:{}$\mu$m colloidal
silica. The final thickness of the samples for SAXS measurement
was approximately 200\:{}$\mu$m. The crystallographic orientation
of the slices was determined by standard Laue diffraction with the
accuracy better than 1\:{}deg.

Small-angle x-ray scattering (SAXS) experiments have been carried
out at the beamline 15-ID at APS, Argonne National Laboratory
(USA). We used the photon energy of 25\:{}keV, the width of the
primary x-ray beam was set to $150 \times 150$\:{}$\mu$m. The
scattered beam was detected by a large two-dimensional detector
with $2048\times 2048$ pixels. One measurement took 60\:{}s and we
took five SAXS pictures for different areas of each sample. The
samples were mounted on a goniometer head allowing for a precise
alignment of the crystallographic axes of the sample lattice with
respect to the primary x-ray beam. For each ageing temperature we
obtained the SAXS patterns in three directions $[001]_\beta,\
[110]_\beta$ and $[111]_\beta$ of the primary beam with respect to
the $\beta$-Ti lattice. The SAXS data were calibrated by a
standard procedure \cite{Ilavsky2009}, using Ag-behenate and
glassy carbon samples for angular and intensity calibrations,
respectively. After the calibration, the scattered intensities
were expressed as photon fluxes scattered from unit sample volume
into unit solid angle of one sr normalized to unit flux density of
the primary radiation and corrected to the sample absorption.

\begin{figure}
\includegraphics[width=12cm]{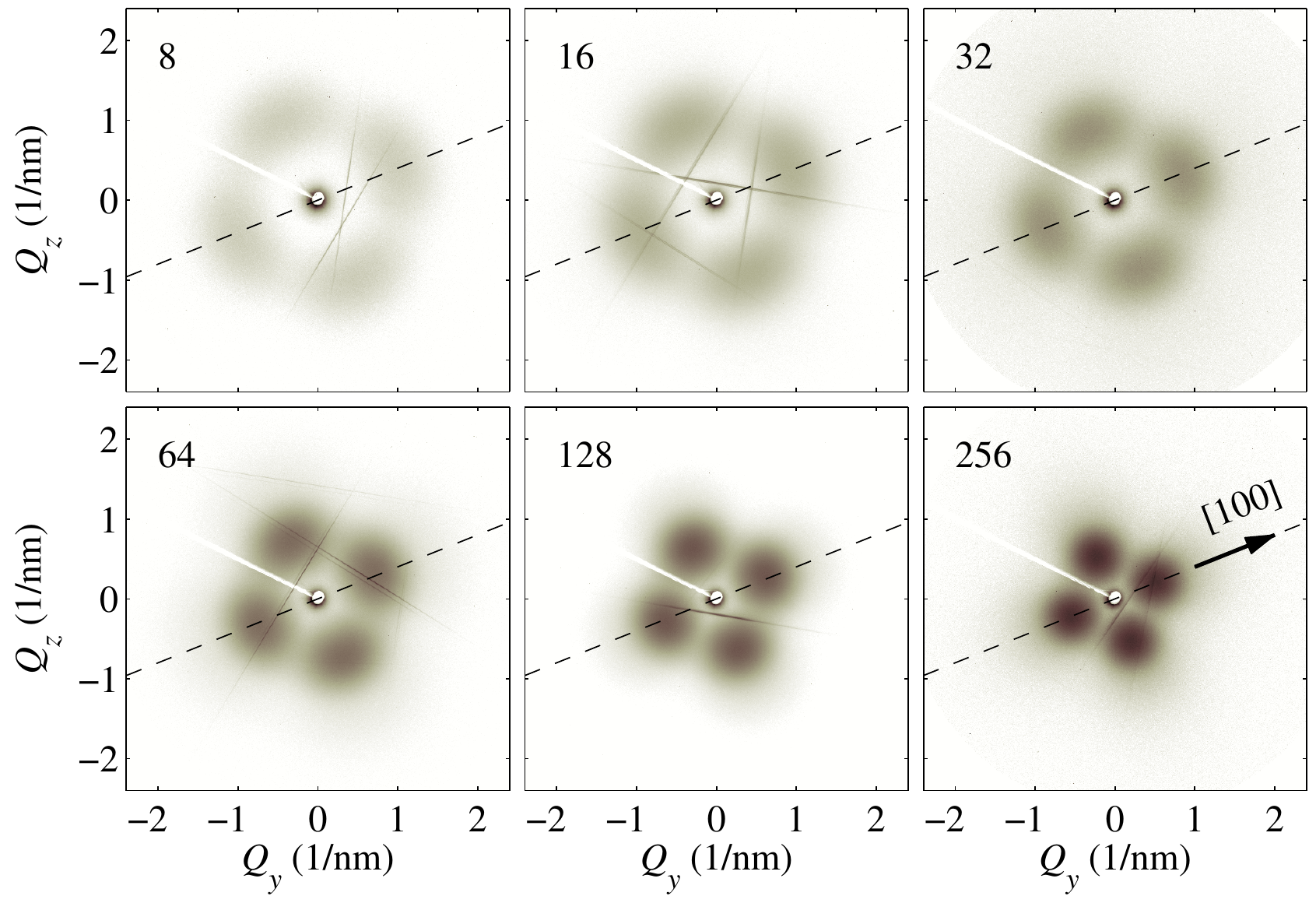}
\caption{The SAXS maps of samples after ageing at 300\oC{}; the
time of ageing in hours is indicated in the upper left corners of
the panels. The measurements have been carried out in the
$[001]_\beta$-orientation of the primary x-ray beam. The line
scans plotted in Figure~\ref{f4} were extracted from these maps
along the dashed lines.} \label{f1}
\end{figure}

\begin{figure}
\includegraphics[width=12cm]{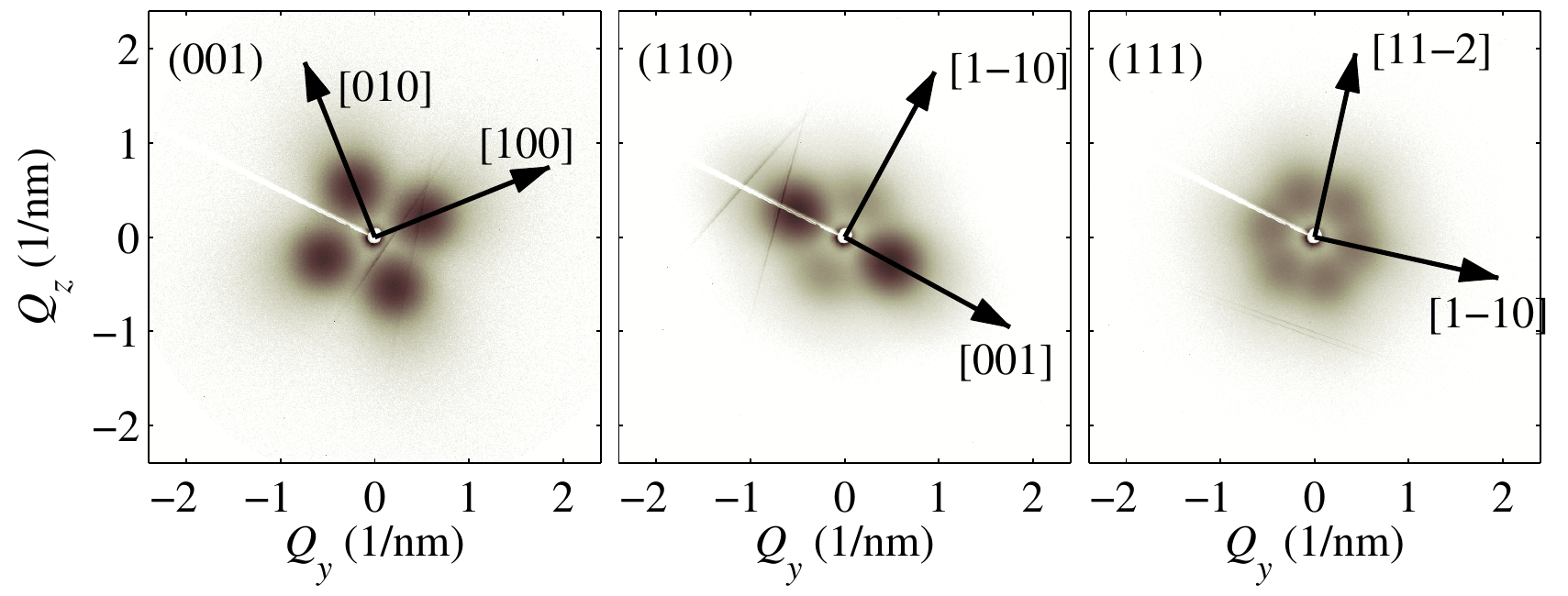}
\caption{The SAXS maps of the sample after 300\oC{}/256\:{}h
ageing measured in three orientations of the primary x-ray beam.
The arrows denote the crystallographic directions in the $\beta$
host lattice determined by the Laue method.} \label{f2}
\end{figure}

In figure~\ref{f1} we plotted the SAXS intensity maps of the
samples after various ageing times (ageing temperature 300\oC{})
measured in the orientation $[001]_\beta$ of the primary x-ray
beam. The arrow denotes the orientation of the $[100]_\beta$
direction in the host lattice determined by independent x-ray
diffraction (the Laue method). The maps clearly exhibit side
maxima in directions $[100]_\beta,\ [010]_\beta$ indicating that
the particles are arranged in a disordered three-dimensional cubic
array with the axes along $\langle 100 \rangle_\beta$. With
increasing ageing time the side maxima move closer to the origin
and they become stronger and narrower; this development can be
explained by increasing mean distance $L_0$ of the neighboring
particles and increasing particle size. Figure \ref{f2} shows the
SAXS maps of the sample after 300\oC{}/256\:{}h ageing measured
for the three crystallographic orientations of the primary beam;
the positions of the side maxima again support the hypothesis of
the ordering of the particles along $\langle 100\rangle_\beta$
axes.

\begin{figure}
\includegraphics[width=12cm]{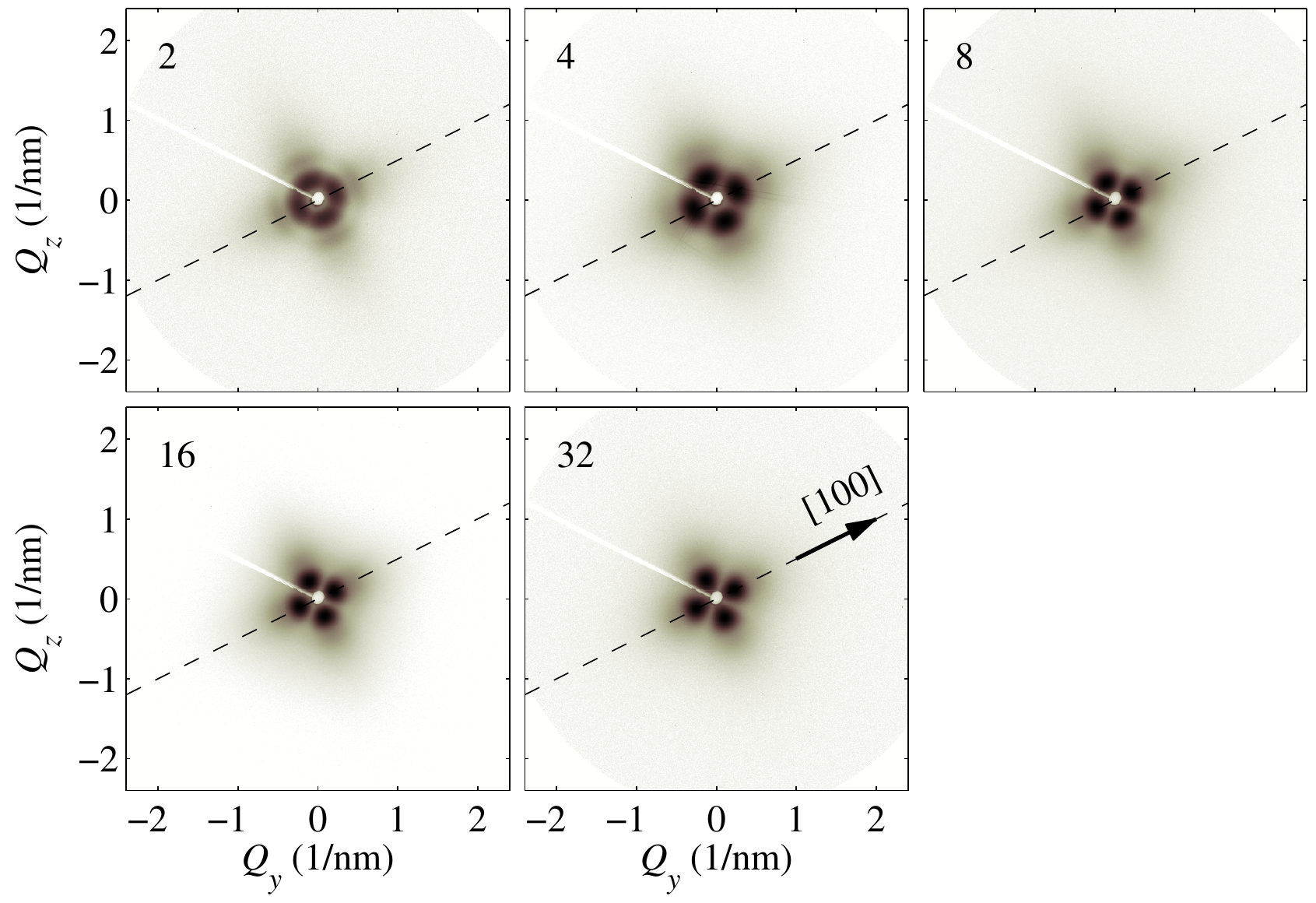}
\caption{The same situation as in Figure~\ref{f1}, ageing
temperature 370\oC{}.} \label{f3}
\end{figure}

The SAXS intensity maps of samples aged at 335\oC{} are quite
similar to Fig.~\ref{f1}, therefore, we do not present them here.
In Fig.~\ref{f3} we show the maps of the samples aged at 370\oC{},
since their appearance obviously differs from figure~\ref{f1}.

Theoretical description of the
small-angle x-ray scattering from an ordered three-dimensional
array of particles along with numerical simulation and fitting to
the experimental data will be described in the next section.

\section{Short-range-order model of the ordering of particles}
\label{SRO}

In this section we present a model for the simulation and fitting
of the SAXS data. The model is purely phenomenological, but its
simplicity makes it possible to fit numerically the measured data.
A more physically substantiated simulation approach will be
presented in Sect. \ref{drivingforce}.

The signal measured in a small-angle-scattering experiment (SAXS)
in a given pixel of a two-dimensional detector
\beq{1}
J=I_{\rm inc} \frac{\D\sigma}{\D\Omega}  \Delta \Omega
\eeq
is proportional to the intensity of the primary beam $I_{\rm
inc}$, the solid angle $\Delta \Omega$ determined by the angular
aperture of the detector pixel and the differential cross-section
$\D\sigma/\D\Omega$ of the scattering process.

The differential scattering cross-section is simulated using a
standard approach \cite{RenaudSSR2009} including the kinematical
approximation (i.e. neglecting multiple scattering from the
particles) and the far-field limit. The explicit formula for the
differential scattering cross-section reads:
\beq{2}
\frac{\D\sigma}{\D\Omega} = \frac{K^4}{4\pi^2} |\Delta n|^2
\E^{-\mu T} \left\langle \sum_n \sum_m \Omega_n^{\rm FT}({\bm
Q})\Omega_m^{{\rm FT}*}({\bm Q}) \E^{-\I {\bm Q}.({\bm r}_n-{\bm
r}_m)}\right\rangle.
\eeq
Here we denoted $K=2\pi/\lambda$, $\mu$ is the linear absorption
coefficient, $T$ is the sample thickness measured along the
primary x-ray beam, and $\Delta n$ is the difference of the
refraction indexes of the particle and the host material (caused
by a slight difference in their chemical compositions). It is
therefore assumed that the refraction index (which is proportional
to the electron density) is homogeneous within the particle and
also the electron density of the host material is assumed
homogeneous. The electron density is determined by chemical
composition (i.e. concentration of impurity atoms Mo, Fe and Al),
as well as by the specific volume per one atom, which is affected
by elastic deformation of the lattice and/or by the presence of
structure defects. $\Omega_n^{\rm FT}(\bm{Q})$ is the Fourier
transformation of the shape function of the $n$-th particle:
\[
\Omega_n^{\rm FT}({\bm Q})=\int \D^3{\bm r} \Omega_n({\bm r})
\E^{-\I {\bm Q}.{\bm r}};
\]
the shape function $\Omega_n({\bm r})$ is unity inside and zero
outside the particle. Further, ${\bm r}_n$ in Eq. (\ref{e2}) is
the position vector of the $n$-th particle, the double sum $\sum_n
\sum_m$ runs over all particles in the irradiated sample volume,
and the averaging $\langle \ \rangle$ is performed over random
positions and sizes of the particles. The scattering vector ${\bm
Q}={\bm K}_f-{\bm K}_i$ is considered in vacuum, since
the refraction correction is unimportant in our transmission
geometry and the absorption effect is included in the absorption
term $\exp(-\mu T)$ in Eq. (\ref{e2}). It is worthy to note that
the measured signal in Eqs. (\ref{e1},\ref{e2}) is proportional to
the structure function defined in Eq. \ref{ea1}:
\beq{2a}
J= I_{\rm inc}  |\Delta n|^2 \frac{K^4}{4\pi^2 Q_{\rm max}^3}
F\left(\frac{{\bm Q}}{Q_{\rm max}}\right)\Delta \Omega,
\eeq
where $F$ is the universal scaling function from Eq. (\ref{ea2}).

In the literature, several models can be found describing a
possible correlation of the particle sizes with their positions
\cite{RenaudSSR2009}. In the following we assume the
local-monodisperse approximation (LMA). Using this approach, the
irradiated sample volume consists of many domains, one domain
contains particles of a given size and given mean distance between
nearest particles. The differential scattering cross-section is
then
\beq{3}
\frac{\D\sigma}{\D\Omega} = \frac{K^4}{4\pi^2} |\Delta n|^2
\left\langle \left|\Omega_{R}^{\rm FT}({\bm Q})\right|^2 G_{R}({\bm
Q})\right\rangle_{\rm sizes},
\eeq
here we have denoted
\beq{4}
G_{R}({\bm Q})=\left\langle \sum_n \sum_m \E^{-\I {\bm Q}.({\bm
r}_n-{\bm r}_m)}\right\rangle_{\rm positions}
\eeq
the correlation function of the positions of the particles with a
given radius $R$. In the following we omit the subscript $R$ for
simplicity.
\begin{figure}
\includegraphics[width=12cm]{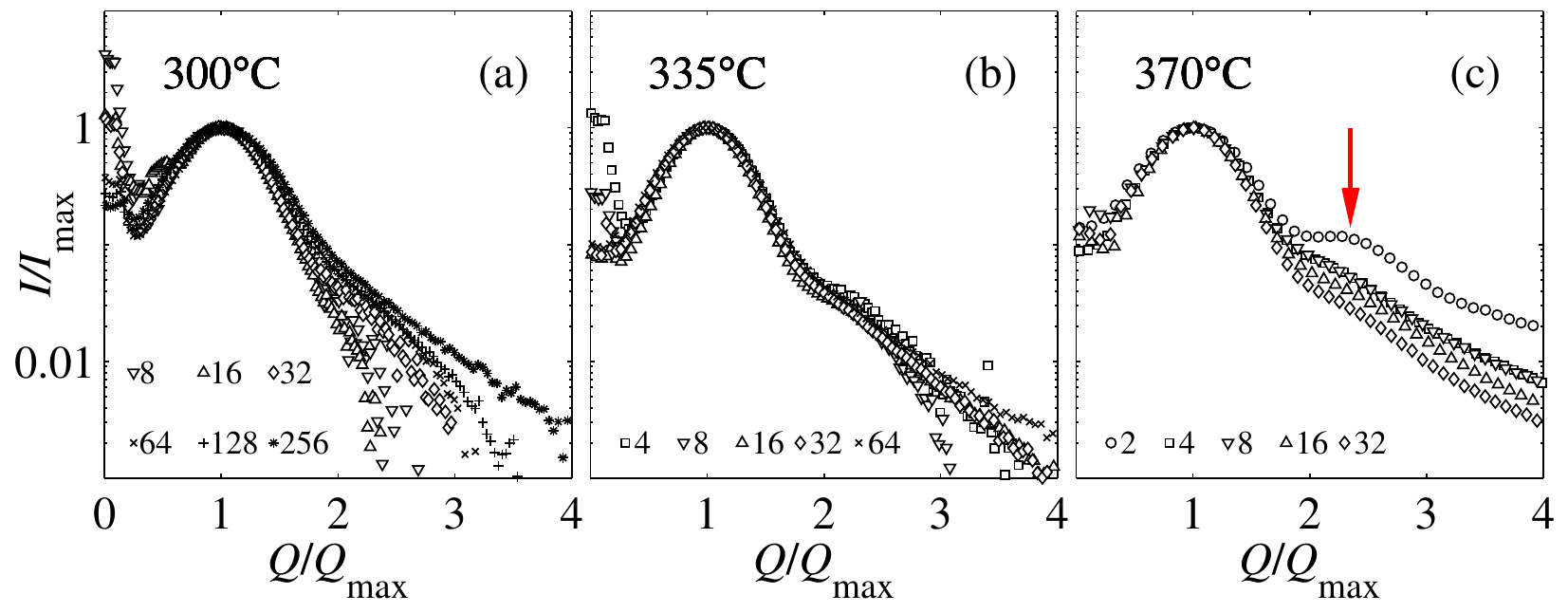}
\caption{Line scans across the satellite maxima measured on
samples after ageing at 300\oC{} (a), 335\oC{} (b), and 370\oC{}
(c). The scans are normalized to the same height and position of
the first satellite maximum. The red arrow in panel (c) denotes
the position of the secondary satellite maximum.} \label{f5}
\end{figure}

Since the position of a particle is only affected by the positions
of the particles in few nearest coordination shells, the ordering
of the particles can be described by a short-range order model
(SRO). In this model we assume that the distances of a given
particle from their neighbors are random with a given statistical
distribution. As the $\omega$ particles create a cubic array, the
three-dimensional correlation function $G({\bm Q})$ can be
expressed as a direct product of three one-dimensional correlation
functions as shown by Eads et~al.~\cite{EadsACA2001}. The
one-dimensional correlation function $G^{(1)}({\bm Q})$ can be
calculated directly \cite{EadsACA2001,BuljanACA2012}:
\beq{5}
G^{(1)}({\bm Q})=N\left\{1+2{\rm Re}\left[ \frac{\xi({\bm
Q})}{1-\xi({\bm Q})} \left(N-\frac{(\xi({\bm Q}))^N-1}{\xi({\bm
Q})-1}\right)\right]\right\}.
\eeq
Here we denoted $N$ the number of coherently irradiated particles
in one dimension and
\[
\xi({\bm Q})=\left\langle \E^{-\I {\bm Q}.{\bm L}} \right\rangle,
\]
where ${\bm L}$ is the random vector connecting the actual centers
of neighboring particles lying in the same one-dimensional chain.
In the following we assume that $N$ is very large and we use the
limiting expression for $G^{(1)}({\bm Q})$:
\[
G^{(1)}({\bm Q}) \rightarrow N\left[1+2{\rm Re}\left(
\frac{\xi({\bm Q})}{1-\xi({\bm Q})}\right)\right].
\]

The three-dimensional correlation function $G({\bm Q})$ can be
expressed as a product of three one-dimensional correlation
functions along the orthogonal axes parallel to the axes of the
cubic array of particles. Since the coordinate $Q$ of the
intensity line scans in Fig.~\ref{f5} is parallel to the array
axis and the other two coordinates along the scans are
approximately zero (we neglect the curvature of the Ewald sphere),
the correlation function used for the simulation of the line scans
is
\beq{5a}
G(Q,0,0)=N_\parallel N_\bot^2 \left(\frac{\sigma_L}{L_0}\right)^4
\left[1+2{\rm Re}\left( \frac{\xi({\bm Q})}{1-\xi({\bm
Q})}\right)\right],
\eeq
where we used the limiting value $N(\sigma_L/L_0)^2$ of the
one-dimensional correlation function for the zero argument.
$N_\parallel$ and $N_\bot$ denote the numbers of the coherently
irradiated particles in the directions parallel and perpendicular
to the primary beam, respectively. $L_0$ is the mean
inter-particle distance and $\sigma_L$ is its root-mean square
(rms) deviation. The total number of the coherently irradiated
particles can be expressed by the irradiated sample volume $V$:
\[
N_\parallel N_\bot^2=\frac{V}{L_0}.
\]

In the simulation we assume that the particles are spherical, and
calculating the function $\xi({\bm Q})$, we apply the condition
that neighboring particles must not intersect. Therefore, in the
averaging over all possible ${\bm L}$'s we excluded the case
$|{\bm L}|<2R$, where $R$ is a random particle radius (assumed
fixed in the averaging over ${\bm L}$'s). Therefore, we assumed a
truncated normal distribution of the random vectors ${\bm L}$ with
the mean value ${L}_0 =\langle |{\bm L}| \rangle$ and rms
deviation $\sigma_L$. Further, we assumed the Gamma distribution
of the particle radii $R$ with the mean value $R_0$ and the rms
dispersion $\sigma_R$; for each radius we considered the
correlation function calculated by Eq. (\ref{e5}) assuming the
mean particle distance $L_0$ proportional to the actual value of
$R$: $L_0=\zeta R$. The averaging in Eq. (\ref{e3}) is then
performed numerically by integrating over $R$-values, keeping
$\zeta$ constant. It can be proved by a direct calculation that
the short-range order model presented here obeys the scaling law
in Eq. (\ref{ea2}), if the mean values $R_0$ and $L_0$ are
proportional ($L_0=\zeta R_0$) and the \emph{relative} rms
deviations are constant, i.e. $\sigma_L \sim L_0$ and $\sigma_R
\sim R_0$.

We have checked the validity of the scaling law in Eq. (\ref{ea2})
by rescaling the line scan to the same position and height of the
satellite maxima, the results are displayed in figure \ref{f5}.
From the figure it follows that at the two lower ageing
temperatures the line scans are scaled according to Eq.
(\ref{ea2}) (panels a and b). The large spread of intensity values
for higher ${\bm Q}$ in Fig.~\ref{f5} is due to high noise due to
background subtraction. However, obvious deviations from the
scaling behavior can be observed for the highest ageing
temperature of 370\oC{}, see Fig.~\ref{f5}(c). In particular, the
line scan of the sample after 2\:{}h ageing exhibits secondary
side maximum (denoted by arrow), which indicates a better ordering
of the particle positions and/or smaller rms deviation of the
particle sizes. We do not observe this secondary maximum for
longer ageing times for the other temperatures. The secondary
maximum gradually disappears during annealing.
\begin{figure}
\includegraphics[width=12cm]{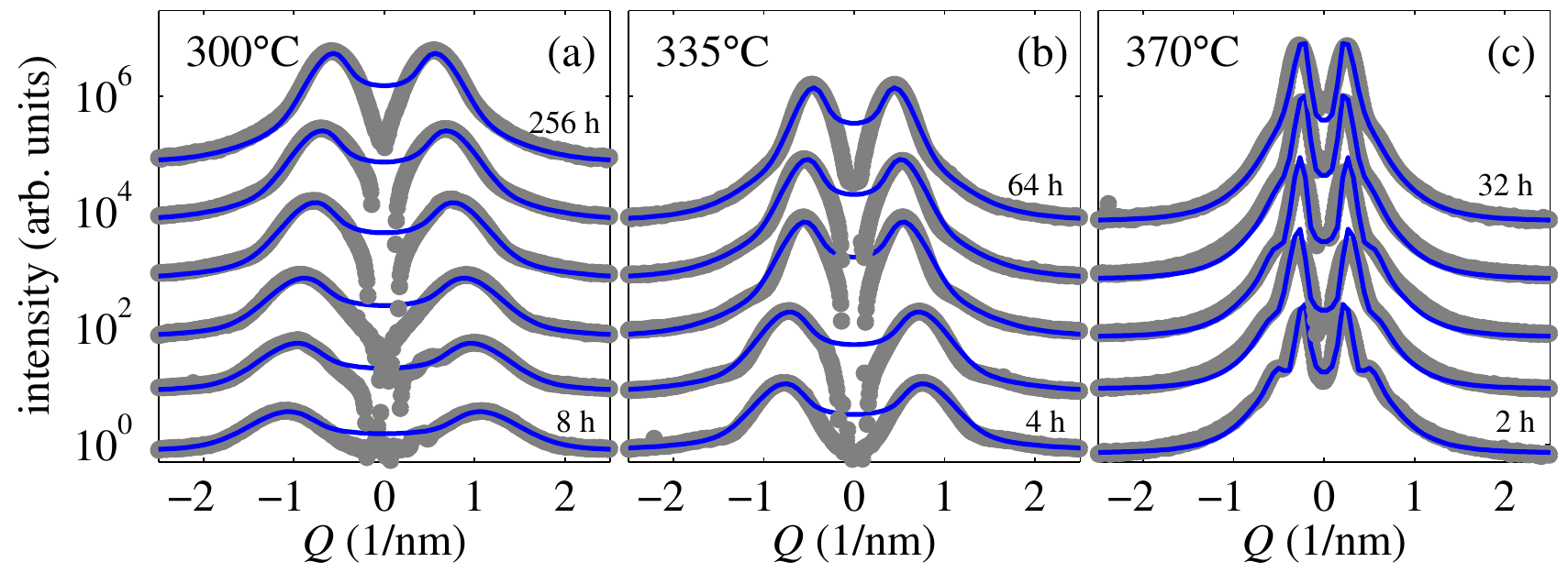}
\caption{The line scans extracted from the SAXS maps (grey dots)
for samples aged at 300\oC{} (a), 335\oC{} (b) and 370\oC{} (c)
and their fits by the SRO model (lines). The numbers denote the
ageing time, the scans are shifted vertically for clarity.}
\label{f4}
\end{figure}

From the SAXS maps we extracted line scans along the $[100]_\beta$
direction (dashed lines in~figures~\ref{f1} and~\ref{f3}) and
fitted them with SRO model. Formulas in Eqs.
(\ref{e3}, \ref{e5}, \ref{e5a}) yield absolute flux densities
therefore from the fit we were able to determine the contrast
$|\Delta n|$ in the refraction indexes of the particle material and
the host phase, appearing in the multiplicative pre-factor.
Nevertheless, in order to determine both $|\Delta n|$ and the mean
particle sizes, we had to assume that the mean inter-particle
distance $L_0$ and the mean radius $R_0$ are proportional, i.e.,
$L_0=\zeta R_0$, and the proportional factor $\zeta$ is the same
in all samples in the same ageing series.

\begin{figure}
\includegraphics[width=7cm]{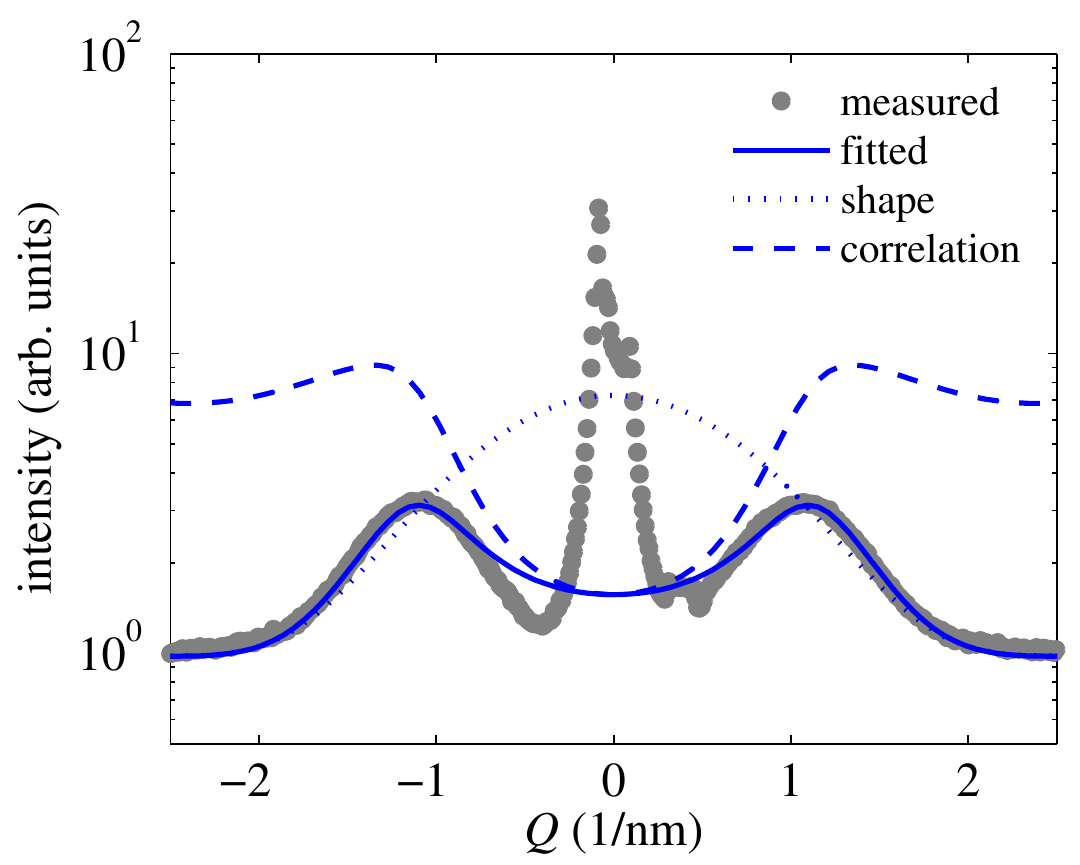}
\caption{The line scan extracted from the SAXS map of sample after
300\oC{}/8\:{}h ageing (points) and its fit by the SRO model (full
line). The contributions of the particle shape and correlation to
the simulated line scan are displayed by dotted and dashed lines,
respectively.} \label{f6}
\end{figure}

Figure \ref{f4} compares the measured (grey points) and fitted
(lines) line scans. It is obvious that the agreement of the theory
with experimental data is quite good. Figure \ref{f6} shows the
measured and fitted line scans of the sample after 300\oC{}/8\:{}h
ageing in more detail; we plotted by dotted and dashed lines the
contributions of the particle shape (function $\langle \left|
\Omega^{\rm FT}({\bm Q}) \right|^2\rangle$) and the correlation
function $G({\bm Q})$ of the particle positions, respectively.
From the figure it is obvious that the shape factor slightly
shifts the side maxima at $Q_{\rm max}$ towards smaller $|Q|$ so
that it would be misleading to determine the mean particle
distance $L_0$ just from the formula $L_0=2\pi/Q_{\rm max}$.

\begin{figure}
\includegraphics[width=12cm]{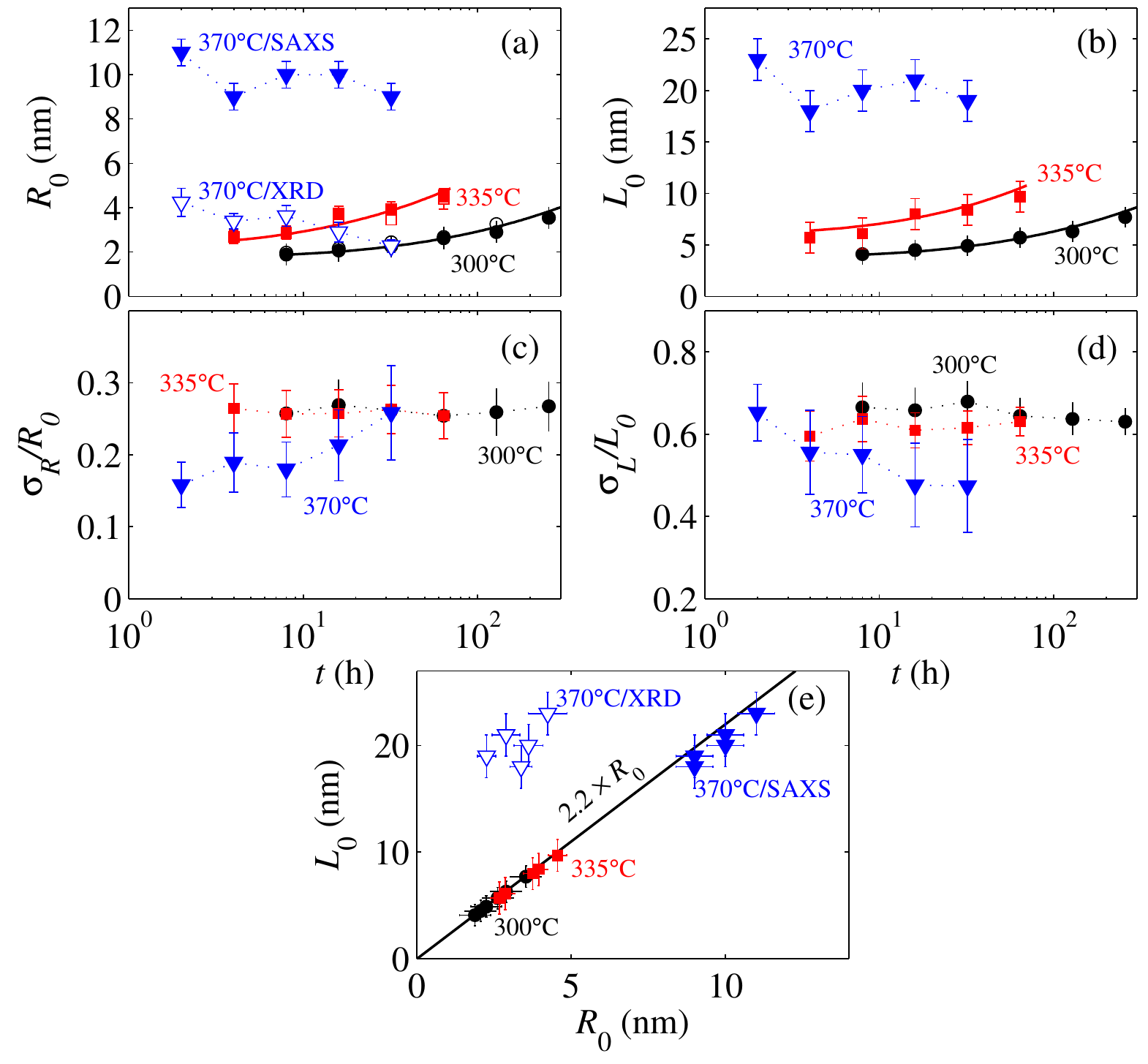}
\caption{Parameters of the particles determined from the fit of
the line scans to the short-range order model. See the text for a
detailed description. The full lines in panels (a) and (b) are the
graphs of fitted functions $(A+Bt)^{1/3}$ ($A$ and $B$ are
suitable constants), following from the LSW theory. The straight
line in (e) represents the dependence $L_0=2.2 R_0$.} \label{f7}
\end{figure}

Parameters of the particle ordering determined from the fits of
the line scans are summarized in Fig.~\ref{f7}. In panel (a) we
plotted the time-dependence of the mean particle radius $R$
determined from the SAXS data. In this panel, we compare these
radii with the particle radii determined by XRD using the method
described in our previous paper \cite{SmilauerovaActaMat2013}. For
the sample series aged at 300\oC{} and 335\oC{} both radii
coincide within the error limits and their time-dependence roughly
agrees with the prediction of the LSW theory $[R_0(t)]^3 \sim A +B
t$ ($A$ and $B$ are constants). The third series aged at 370\oC{}
behaves in a different way. With increasing ageing time the
particle radii determined from XRD decrease, however, the error
bars of these radii are larger than for the other ageing
temperatures. In order to obtain a reasonably good fit of the data
from the third sample series, we had to fit the parameters $L_0$
and $R_0$ \emph{independently}, not considering the
proportionality factor $\zeta$. This fact makes the fitting
results less reliable than for the other two series, however, it
is obvious that the particles sizes determined from SAXS are
larger than those from XRD. Figure \ref{f7}(b) displays the time
dependence of the mean particle distance $L_0$ determined from
SAXS. Again, for sample series at 300\oC{} and 335\oC{} $L_0$
increases with the ageing time and follows the polynomial formula
$[L_0(t)]^3 \sim A' +B' t$ following from the LSW theory. For the
highest ageing temperature, no distinct evolution of the $L_0$
values during ageing can be established.

In figure \ref{f7}(c) we have plotted the time dependence of the
relative rms deviation $\sigma_R/R_0$ of the particle radii. In
the first two sample series aged at 300\oC{} and 335\oC{} the relative
rms deviation does not change significantly during ageing, while
at the highest ageing temperature of 370\oC{} we observe a distinct
increase of this value, i.e. the width of the size distribution of
the particles increases during ageing. The relevance of this
result is somewhat limited by the fact that the fit of the SAXS
data of the last sample series is less reliable than for the other
two temperatures (see the discussion in Sect. \ref{discussion}), however the qualitative tendency is obvious.

Figure~\ref{f7}(d) shows the time dependence of the relative rms
deviation $\sigma_L/L_0$ of the inter-particle distances. At
300\oC{} and 335\oC{} these rms deviations remain nearly constant,
while at 370\oC{} they slightly decrease with the ageing time,
however the errors of these parameters are quite large.
Therefore, the second-order maxima in the line
scans depicted in Fig.~\ref{f5}(c) can be ascribed to the form-factor of a single particle and not to the correlation function of the particle positions. Finally, in Fig.~\ref{f7}(e)
we demonstrate that the mean inter-particle distance $L_0$ scales linearly with the mean
particle radius $R_0$ determined from SAXS, obeying the approximative formula $L_0
\approx 2.2 \times R_0$.

\begin{figure}
\includegraphics[width=7cm]{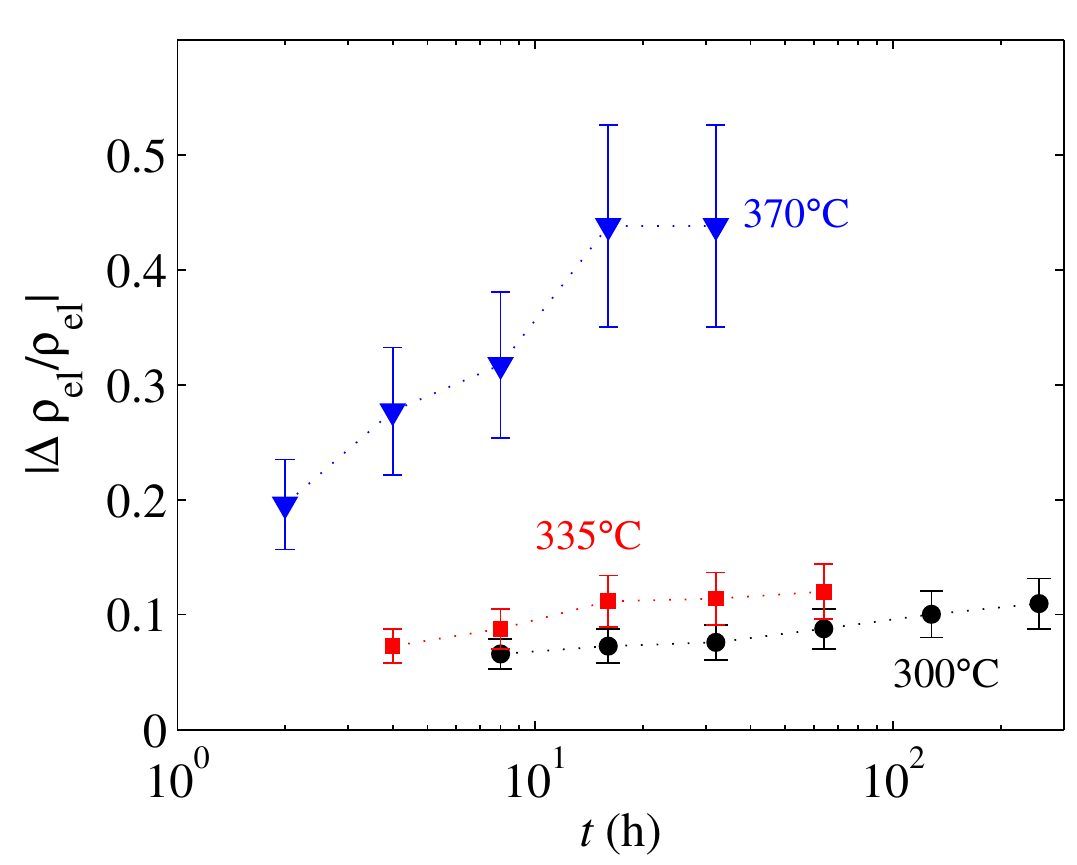}
\caption{The difference of the electron density $\Delta \rho_{\rm el}$ of the $\omega$ particles and the $\beta$-Ti matrix relatively to the electron density of the matrix $\rho_{\rm el}$ \emph{vs} ageing time.} \label{f8}
\end{figure}

Finally, from the fits we determined the contrast of the
refraction index $|\Delta n|$. The difference of the refraction indexes of the particle material and the matrix is proportional to the difference $\Delta \rho_{\rm el}$ in the electron densities:
\[
\Delta n = - \frac{\lambda^2 r_{\rm el}}{2\pi} \Delta \rho_{\rm el},
\]
where $\lambda$ is the x-ray wavelength, $r_{\rm el} \approx 2.818$~\AA$^{-1}$ is the classical electron radius, and we neglected the dispersion corrections. In Fig. \ref{f8} we plotted the contrast of the electron densities relatively to the electron density of the nominal Ti alloy (according to Tab. \ref{t1}) as functions of the ageing time.
In all sample series, the contrast $|\Delta \rho_{\rm el}|$ increases with ageing time. During the ageing at 300\oC{} and 335\oC{}, the contrast values are smaller or around 10\% of the nominal value, these changes can be
explained by the changes in the chemical composition of the
particles by several at. \% of Mo, Fe, and/or Al. Of course, one single
value of $|\Delta \rho_{\rm el}|$ for a given sample does not allow to
determine complete chemical composition of the particles. For the
highest ageing temperature of 370\oC{} the contrast values following
from the fit are much larger and do not correspond to any physically relevant value. This result will be discussed in Sect. \ref{discussion}.

\section{Driving force of the ordering}
\label{drivingforce}

In the previous section we demonstrated that the ordering of the
particles agrees well with the short-range order model. In this
section we show that the driving force of the ordering can be
attributed to the minimization of the elastic interaction energy
of the particles.

We have verified in our previous paper
\cite{SmilauerovaActaMat2013} that the crystal lattice around a
particle is elastically deformed, the reason of the deformation is
a difference between the actual lattice parameters $a_\omega,\
c_\omega$ of the $\omega$ lattice of the particle and their ideal
values $a_\omega^{(\rm id)}, c_\omega^{(\rm id)}$ following from
the topotaxy relation of the $\beta$ and $\omega$ lattices
\cite{silcock1958x,FratzlActaMet1991}. Most likely, this lattice
mismatch is caused by a difference in the chemical composition;
during the formation and growth of $\omega$ particles the
$\beta$-stabilizing impurities (Mo and Fe in our case) are
expelled from the particle. Since the $\beta$-Ti matrix is highly
elastically anisotropic, the local deformation field around a
particle is anisotropic, too. The interaction energy of a particle
pair is given by the formula
\cite{EshelbyPRS1957,ArdellACM1966,ShneckPhM1992}
\beq{6}
E_{\rm int}=-\int_{\Omega^{\rm (B)}} \D^3 {\bm r} \sigma^{\rm
(A)}_{jk}({\bm r}) \epsilon_{0jk}^{\rm (B)}({\bm r}), \ j,k=x,y,z.
\eeq
The integral in this formula is calculated over the volume
$\Omega^{\rm (B)}$ of particle B, $\hat{\sigma}^{\rm (A)}({\bm
r})$ is the stress tensor in the matrix in the points belonging to
$\Omega^{\rm (B)}$, caused by another particle A, and
$\hat{\epsilon}_0^{\rm (B)}$ is the mismatch of the lattice of
particle B with respect to the host lattice. Using the mismatch
values
\[
f_a=(a_\omega-a_\omega^{\mathrm{(id)}})/a_\omega^{\mathrm{(id)}},\
f_c=(c_\omega-c_\omega^{\mathrm{(id)}})/c_\omega^{\mathrm{(id)}}
\]
defined in our previous paper and using the coordinate axes across
and along the $c$-axis $[0001]_\omega$ of the hexagonal $\omega$
lattice, the matrix $\hat{\epsilon}_0^{\rm (B)}$ has the form
\beq{7}
\hat{\epsilon}_0^{\rm (B)}=\left( \begin{array}{ccc}
f_a & 0 & 0\\
0 & f_a & 0\\
0 & 0 & f_c
\end{array}
\right).
\eeq
The stress tensor $\hat{\sigma}^{\rm (A)}({\bm r})$ caused by the
particle A was calculated taking into account the elastic
anisotropy of the host lattice and the mismatch matrix
$\hat{\epsilon}_0^{\rm (A)}$ analogous to that in Eq. (\ref{e7})
using the continuum elasticity approach briefly described in the
Appendix of our previous paper \cite{SmilauerovaActaMat2013}.

\begin{figure}
\includegraphics[width=7cm]{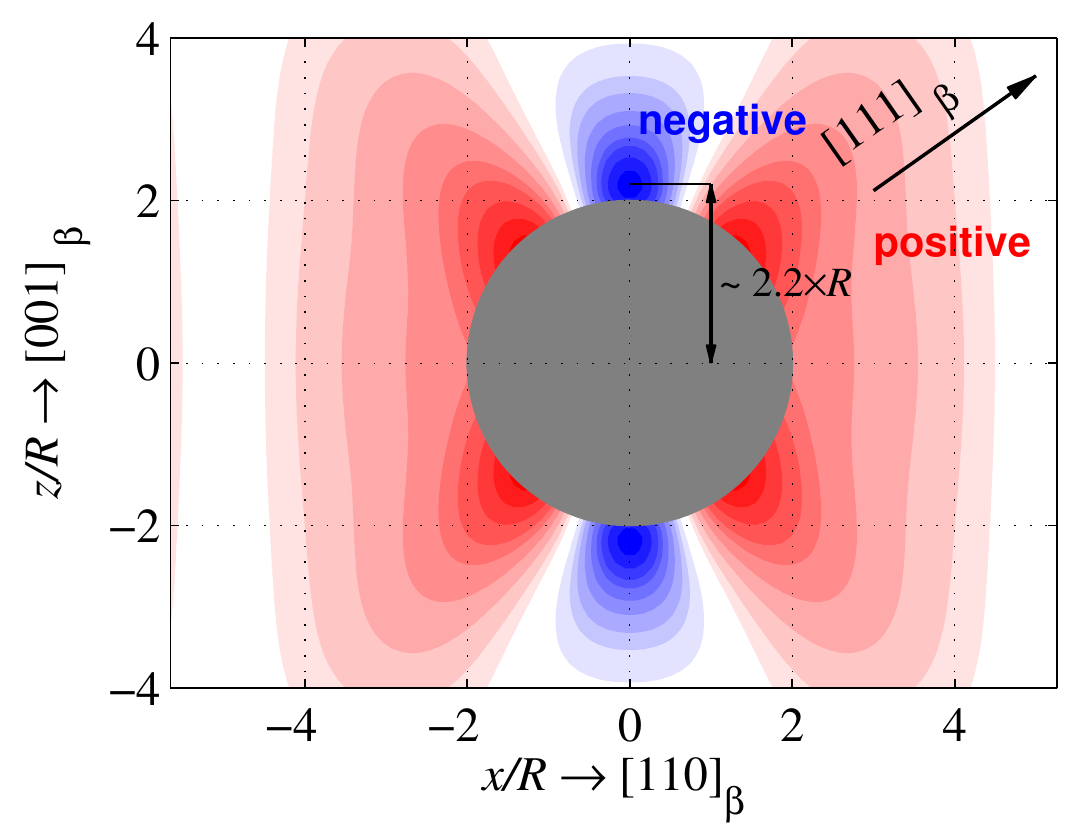}
\caption{Dependence of the interaction energy of a particle pair
on the relative position of the particles. The simulation was
performed for spherical particles with the radius of 3.6\:{}nm and
with the mismatch values $f_a=0.002,\ f_c=0.01$, taking into
account all possible orientations of the hexagonal
$[0001]_\omega$-axes in both particles. The grey area denotes the
region, where the particles intersect. The contour step is
0.1\:{}eV.} \label{f9}
\end{figure}

Assuming the typical mismatch values $f_a=0.002$ and $f_c=0.01$
found in \cite{SmilauerovaActaMat2013} and the particle radius
$R=3$\:{}nm we calculated the dependence of the interaction energy
on the relative position ${\bm r}$ of particles (Fig.~\ref{f9}) in
the $(1\bar{1}0)_\beta$ plane in the cubic $\beta$-Ti lattice. In
this figure, the center of one particle is in the graph origin;
since Eq. (\ref{e6}) is valid only for non-intersecting particles,
the excluded region $|{\bm r}| \le 2R$ is shaded (the grey area).
The simulations were performed for all 16 combinations of the
orientations of the hexagonal $c$-axes $[0001]_\omega$ of
particles A and B with respect to the cubic $\beta$ lattice, the
interaction energy plotted in this figure is averaged over all
orientations.

The figure clearly indicates that minima of the interaction energy
occur in six equivalent directions $\langle 100 \rangle_\beta$
from the particle center in the distance of about $2.2\times R$.
On the other hand, maxima of the interaction energy occur along
eight equivalent directions $\langle 111 \rangle_\beta$. Numerical
simulations demonstrated that the anisotropy in the distribution
of $E_{\rm int}$ is determined entirely by the elastic anisotropy
of the host lattice and it is only very slightly affected by the
anisotropy of the mismatch according to Eq. (\ref{e7}).

The $\langle 100\rangle_\beta$ directions in which the minima of
$E_{\rm int}$ occur agree with the orientations of the basis
vectors of the disordered array of particles determined from the
SAXS data in the previous section. Therefore the anisotropy in the
distribution of interaction energy indicates that the interaction
energy plays a role in the self-ordering mechanism of the
particles. In order to support this hypothesis we performed a
simple Monte-Carlo (MC) simulation of the distribution of particles.
MC simulations are widely used in the simulation of x-ray diffuse
scattering and small-angle scattering. Our MC simulation program
is similar to the MC simulation program for small-angle neutron
scattering (SANS) \cite{StrunzJAC2003}, however, it takes into
account elastic interaction between the particles.

The simulation procedure consists of the following steps:
\begin{enumerate}
\item
we determine randomly the particle radius $R$ using a random number generator, assuming the Gamma distribution of the radii with the mean value $R_0$ and order $m_R$,
\item
we choose randomly the position of the first particle in the
simulation cube $D\times D \times D$,
\item
we choose randomly the position of a next particle and one of four
possible orientations of its hexagonal $[0001]_\omega$-axis,
\item
we calculate the total interaction energy $E_{\rm int}$ of this
particle with other particles seated in the previous steps,
\item
we generate a random number $p \in [0,1]$ and we settle the
particle in the position chosen in the previous step if $p<
K\exp[-E_{\rm int}/(k_B T)]$,
\item
we repeat items 3-5 $N$ times, where $N$ is the number of attempts
to place a particle,
\item
we repeat items 1-6 $M$ times, where $M$ is the number of
simulation cubes,
\item
we calculate the scattered intensity using the formula
\beq{8}
J({\bm Q})={\rm const.} \sum_{k=1}^M \left|\Omega^{\rm
FT}_{R_k}({\bm Q})\sum_{n=1}^{N_k} \E^{-\I {\bm Q}.{\bm
r}_n^{(k)}} \right|^2,
\eeq
where ${\bm r}_n^{(k)}$ are the particle position vectors
generated in items 1-5, and $N_k \le N$ is the actual number of
settled particle for given $k$-th simulation cube.
\end{enumerate}

Therefore, the simulation procedure has the following parameters:
$R_0$ is the mean radius of the particles, $m_R=(R_0/\sigma_R)^2$ is the order of the Gamma distribution of the radii. $D$ is the size of the
simulation domain and it is comparable to the coherence width
and/or length of the primary x-ray beam; we took $D=50$\:{}nm.
The constant $K$ was chosen so that the values $K\exp[-E_{\rm
int}/(k_B T)]$ lie between 0 and 1, i.e. $K\approx\exp[{\rm
min}(E_{\rm int})/(k_B T)]$ and we found that the simulation
results do not depend much on $K$. The simulation temperature $T$
is not directly connected to the ageing temperature and we choose
the value of $T$ to obtain the best match of the simulation
results to the experimental data, namely $k_BT=0.5$\:{}eV. The
number $N$ of the trials to set the particle positions was chosen
much larger than the expected number of the particles in the
$D\times D\times D$ cube; we used $N=10^6$. The number $M$ of the
simulation cubes is determined by the ratio of the total
irradiated sample volume to the coherently irradiated volume. This
ratio is roughly $10^9$ for our experimental conditions, however
we used $M=10^3$ to keep the calculation time in reasonable
limits. The MC simulation procedure is only qualitative, since it
describes properly neither the microscopic mechanism of the $\beta
\rightarrow \omega$ transition, nor the growth of nucleated
$\omega$ particles.
\begin{figure}
\includegraphics[width=12cm]{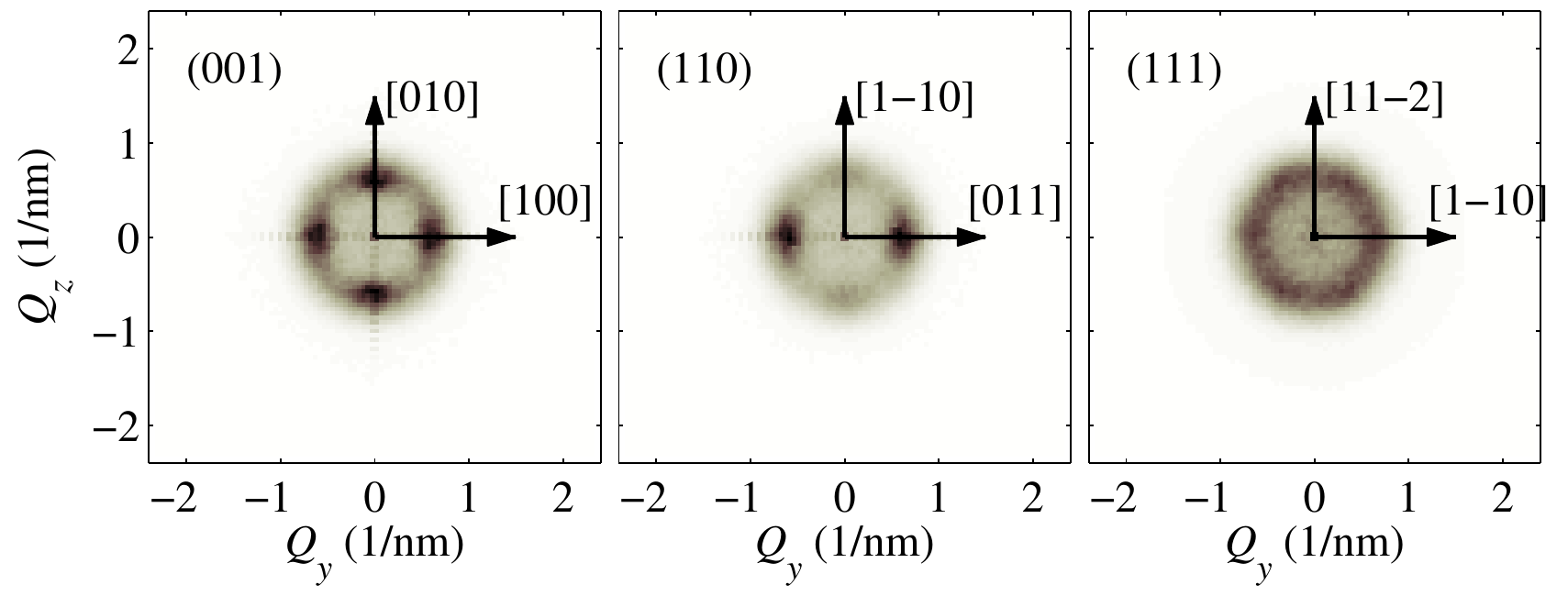}
\caption{SAXS maps simulated in three reciprocal planes $(001),\
(110)$ and $(111)$ by the Monte-Carlo method described in text.}
\label{f10}
\end{figure}

Figure \ref{f10} shows the examples of the simulated SAXS maps in
planes $(001),\ (110)$ and $(111)$ perpendicular to the primary
x-ray beam. In the simulations we took $R_0=3.6$~nm and $m_R=20$.
The maps exhibit distinct side maxima, the positions of which very
well coincide with the maxima in the measured maps in
Fig.~\ref{f2}. The distance of the simulated maxima from the
origin ${\bm Q}=0$ is inversely proportional to the mean radius
$R_0$ of the particles; from the simulation we found that the
position $Q_{\rm max}$ of the maximum at the $[100]$ axis obeys
the formula
\beq{9}
\frac{2\pi}{Q_{\rm max}}=L_0=\zeta R_0; \zeta = 2.6 \pm 0.05.
\eeq
The factor $\zeta \approx 2.6$ found from the MC simulations is
slightly larger that the proportionality factor 2.2 between the
position of the minimum of the interaction energy of a particle
pair and the particle radius (see Fig.~\ref{f9}). This slight
discrepancy might be caused by the fact that many particles (not
only the nearest ones) contribute to the total interaction energy
of a given particle. Another reason could be the asymmetry of the
statistical distribution of the inter-particle distances stemming
from the fact that the neighboring particles must not penetrate.
However, from the SAXS data $\zeta = 2.2 \pm 0.1$ follows (see
Fig.~\ref{f7}(e)).

\begin{figure}
\includegraphics[width=7cm]{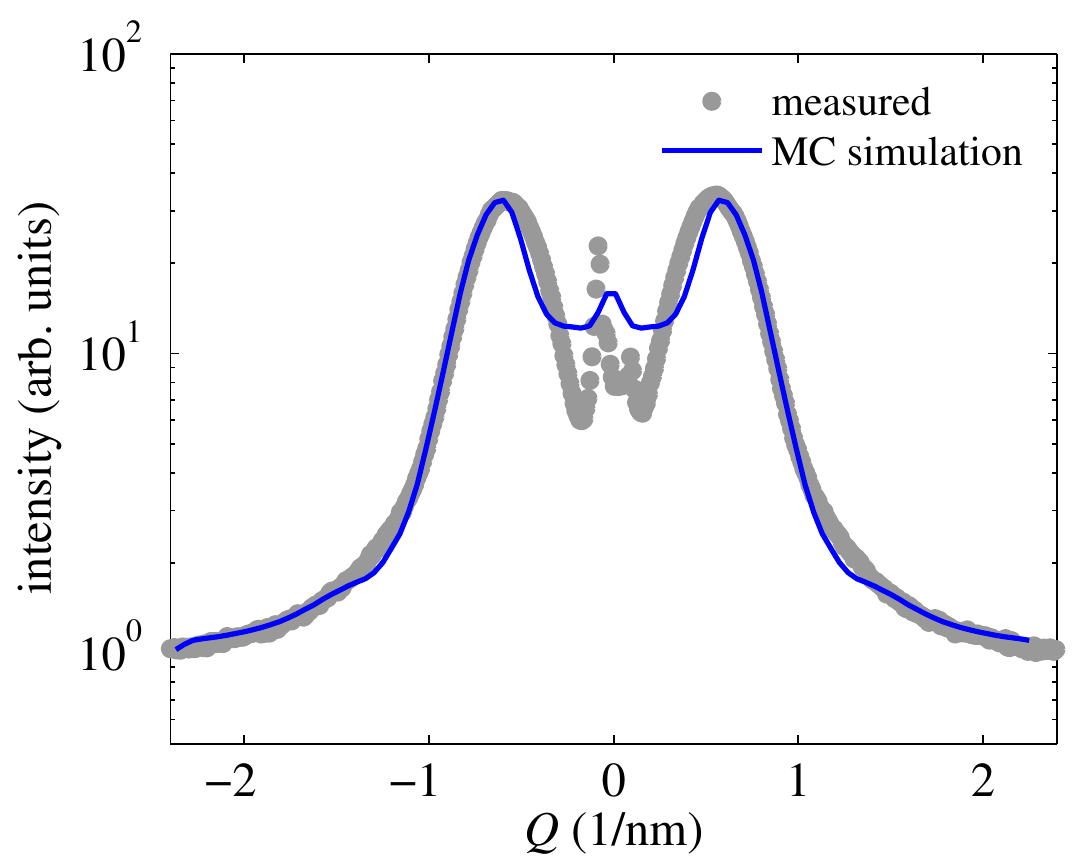}
\caption{Comparison of the line scan extracted from the measured
SAXS map in Figure~\ref{f1} of sample after 300\oC{}/256\:{}h ageing
taken in direction $[001]_\beta$ (points) with the result of the
Monte-Carlo simulations performed for $R_0=3.6$\:{}nm and $m_R=20$
(line).} \label{f11}
\end{figure}

A direct comparison of the measured and MC-simulated line scans is
plotted in figure~\ref{f11}, where we compare the line scans
extracted from the measured SAXS map of sample after
300\oC{}/256\:{}h ageing in figure~\ref{f1} with the MC simulation
performed for $R_0=3.6$\:{}nm and $m_R=20$; the simulated
intensities were multiplied by a suitable constant to obtain the
same heights of the side maxima. The shapes of the intensity
distributions coincide well.

\section{Discussion}
\label{discussion}

The SAXS data were compared with simulations based on a
phenomenological SRO model and we found a reasonably good
agreement (see figure~\ref{f4}). From the fit we determined the
mean particle radius $R_0$ and inter-particle distance $L_0$ and
their dependence on the ageing time $t$. In samples aged at
300\oC{} and 335\oC{} the mean particle radii determined from SAXS
and XRD coincide within the error limits. Furthermore, in
agreement with the LSW model, the radius $R_0$ and the distance
$L_0$ increase roughly as $t^{1/3}$, i.e. the total number $N$ of
particles decreases as $1/t$ in these samples. The same scaling
laws were also demonstrated from the XRD data in our previous
paper, so that both XRD and SAXS data are consistent and they
confirm the validity of the LSW model for the ageing temperatures
300\oC{} and 335\oC{}.

\begin{figure}
\includegraphics[width=12cm]{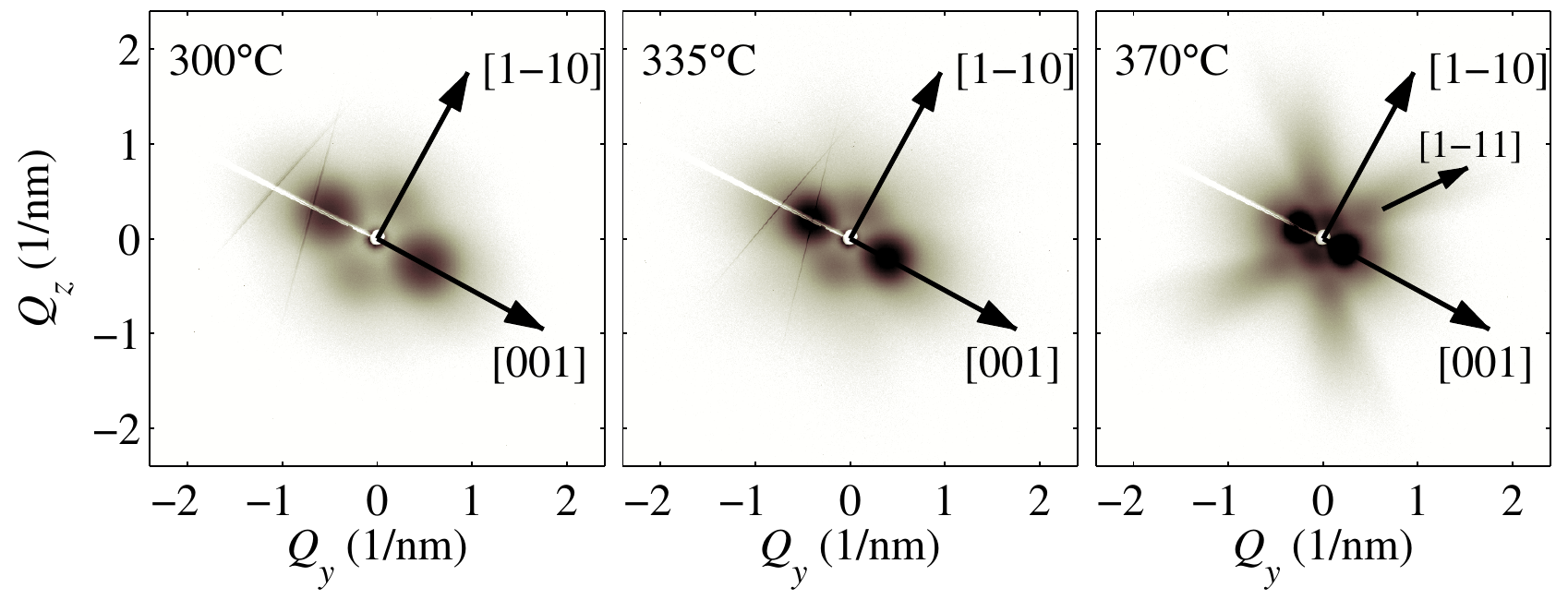}
\caption{The SAXS intensity maps of the last samples of all three
ageing series measured in the $(110)_\beta$ plane. The
$[1\bar{1}1]_\beta$-oriented streak is clearly visible in the map
of the sample aged at the highest temperature.} \label{f12}
\end{figure}

The samples aged at the highest temperature of 370\oC{} behave
differently, namely, the mean particle distance $L_0$ and the mean
radius $R_0$ determined from SAXS remained nearly constant during
ageing, while the XRD-determined particle sizes are much smaller.
The main reason might be that 370\oC{} is temperature sufficient
for $\alpha$ phase particles precipitation. It is well-known that
the $\alpha$ particles have the form of platelets
\cite{FratzlActaMet1991,LangmayrPRB1994} parallel to the
$(0001)_\alpha$ basal planes perpendicular to $\langle 111
\rangle_\beta$ directions. In SAXS, such platelets give rise to
intensity streaks along $\langle 111 \rangle_\beta$; these streaks
should be visible in the SAXS intensity maps in the orientation
$(110)_\beta$. In Fig.~\ref{f12} we compare the SAXS maps of the
last samples of all ageing series 300\oC{}, 335\oC{}, and
370\oC{}. A $[1\bar{1}1]_\beta$-oriented streak is clearly visible
indeed only in the map of the sample aged at the highest
temperature of 370\oC{}. Full description of these streaks and the
evaluation of size of $\alpha$ platelets are beyond the scope of
this paper.

In the structure model used for the fitting of the SAXS data [Fig. \ref{f4}(c)] we did not include the $\alpha$ platelets, which increase the scattered intensity for small $Q$'s. Consequently, the parameters resulting from the fit of this data series are less reliable. This affects mainly the values $\Delta \rho_{\rm el}$ of the contrast of the electron density in Fig. \ref{f8}. The $\Delta \rho_{\rm el}$ values of the 370\oC{} series are strongly overestimated, since the scattered intensity was ascribed only to the $\omega$ particles, and a part of the intensity stems also from the $\alpha$ platelets.

At the highest temperature (370\oC), the size of the $\omega$ particles seen by XRD is smaller than the size detected by SAXS. This temperature may be high enough for the $\omega$ particles to grow quickly at the beginning of ageing and then start to dissolve at longer ageing times (or to transform to the $\alpha$ phase). As the $\omega$ structure disappears, XRD detects smaller size of the $\omega$ particles. On the other hand, SAXS detects inhomogeneities in the electron density (i.e. chemical composition), which may remain the same even after the $\omega$ phase dissolves. However, this hypothesis would need more thorough investigation.

A gradual change in the mean chemical composition of the $\omega$
particles during ageing at 300\oC{} and 335\oC{} is the reason for
the slight increase of the $\Delta \rho_{\rm el}$ values in Fig.~\ref{f8}. The
increase of the chemical contrast during ageing could be ascribed
to a gradual ejection of the $\beta$-stabilizing elements (Mo and
Fe in our case) from the volumes of the $\omega$ particles during
the ageing process. In the 370\oC{} sample series, the $\Delta \rho_{\rm el}$
values are much larger and they cannot be explained by mere
chemical changes. Most likely, aforementioned shell structures are
the reason for these values, however this effect requires further
investigation.

From the SAXS data it also follows that for ageing temperatures of
300\oC{} and 335\oC{} the mean particle distance $L_0$ is
proportional to their mean radius $R_0$. This finding indicates
that a particle-particle interaction is the reason of the
ordering. Nevertheless, the phenomenological SRO model used here
\emph{cannot} explain fully the SAXS data. In this model the
position of a given particle is affected only by the positions of
\emph{neighboring} particles. On the other hand, the
inter-particle interaction mediated by elastic deformation of the
host lattice is long-ranged and the position of a given particle
is therefore affected by more distant particles as well. The SRO
model fails especially between the central peak and lateral
maxima, where the measured intensity exhibits a deeper dip than
the simulated curve for ageing temperatures of 300\oC{} and
335\oC{}. The shape of the intensity distribution in this region can
be affected by the asymmetry of the statistical distribution of
the random vectors ${\bm L}$ connecting neighboring particles.
Another reason of the discrepancy between the measured and
simulated data for small $|Q|$ could be the above-mentioned
core-shell structure of the particles modifying the radial profile
of the refraction index.

From the SAXS data shown above it clearly follows that the
$\omega$ particles are self-ordered in a three-dimensional cubic
array with the axes along $\langle 100 \rangle_\beta$ directions.
This finding differs from the conclusions in Ref.
\cite{PrimaJMSL2002}, where the authors claim that the particle
ordering occurs along directions $\langle 111 \rangle_\beta$. The
authors support this statement by a transmission electron
micrograph (TEM), where only few particles are depicted. The
ordering along three $\langle 100 \rangle_\beta$ directions may in
certain cases appear as $\langle 111 \rangle_\beta$ ordering in
TEM, but the statistical relevance of SAXS data is much higher,
since the number of irradiated particles in a typical SAXS
experiment is several $10^9$, i.e. by many decades larger than in
TEM. The $\langle 100 \rangle_\beta$-oriented ordering of
particles can be explained by the following simple argument. As we
have shown above, the arrangement of the particles is close to a
thermodynamic equilibrium, i.e. the particle positions correspond
to the minima of the interaction energy of particles. As stated by
Shneck et al. \cite{ShneckPhM1992}, the sign of the hydrostatic
stress, i.e. the sign of the trace ${\rm Tr}(\hat\sigma)$ of the
stress tensor, is decisive for the ordering. Namely, if a particle
compresses the surrounding lattice (which is the case of our
samples) and ${\rm Tr}(\hat\sigma)<0$ in the position where a new
particle would appear, then the interaction energy between the
existing particle and another newly formed particle is positive
(repulsive). Indeed, in this case the new particle works against
the stress field of the existing particle and the potential energy
of the particle pair increases. Our finding is also in agreement
with Ref. \cite{FratzlJSM1999}, stating that the particle ordering
occurs along an elastically soft direction, i.e. along $\langle
100\rangle_\beta$ in our case.

We performed a series of Monte-Carlo simulations explaining
qualitatively the ordering mechanism. The positions of the SAXS
maxima and the linear dependence of the mean particle distance on
the size of the particles following from the simulations agrees
well with the SAXS data. However, a detailed comparison of the
experimental data with the simulation results is not possible,
since the simulation model is not fully atomistic. It does not
take into account both the atomistic mechanism of the
$\beta\rightarrow\omega$ transition and the kinetics of the
particle formation and growth.

\section{Summary}
\label{summary}

We have studied the sizes and positions of hexagonal $\omega$ Ti
particles in single crystals of cubic $\beta$-Ti alloy by
small-angle x-ray scattering. We determined the dependence of the
particle size and distance on the ageing time and demonstrated
that the particle growth can be described by the LSW model
\cite{LifshitzJPCS1961,WagnerZEE1961}. We found that the particles
spontaneously order creating a cubic three-dimensional array with
the axes along the cubic axes $\langle 100 \rangle_\beta$ of the
host lattice. The structure of the array can be described by a
phenomenological short-range order model and we demonstrated by a
Monte-Carlo simulation that the driving force of the ordering is
the minimization of the elastic energy of inter-particle
interactions.

\section*{Acknowledgements}

The authors gratefully acknowledge  prof. Henry J. Rack for
helpful comments on phase transformations in Ti alloys and for the
idea of their investigation by the means of SAXS. The work was
supported by the Ministry of Education, Youth and Sports of Czech
Republic (Project LH13005), by the Czech Science Foundation
(Projects P204/11/0785 and 14-08124S), and by the Grant Agency of Charles
University in Prague (Project 106-10/251403). The single-crystal
growth was performed in MLTL (http://mltl.eu/) within the program
of Czech Research Infrastructures (Project No. LM2011025). The
ChemMatCARS Sector 15 of the synchrotron source APS is principally
supported by the National Science Foundation/Department of Energy
under grant number NSF/CHE-0822838. Use of the Advanced Photon
Source, an Office of Science User Facility operated for the U.S.
Department of Energy (DOE) Office of Science by Argonne National
Laboratory, was supported by the U.S. DOE under Contract No.
DE-AC02-06CH11357.

\end{document}